\documentclass{article}
\usepackage{spconf,amsmath,amssymb,graphicx}
\usepackage{booktabs,multirow}
\usepackage[table]{xcolor}
\usepackage{siunitx}
\usepackage{cite}
\usepackage{float}
\usepackage{bbding}
\usepackage{color}
\usepackage{enumitem}
\usepackage{etoolbox}
\robustify\bfseries
\usepackage{balance}
\usepackage{bm}
\usepackage{caption}
\usepackage{subcaption}
\usepackage{circledsteps}
\usepackage{pifont}
\usepackage{arydshln}

\usepackage{makecell}
\usepackage{threeparttable}
\usepackage{flushend}

\usepackage{hyperref}

\makeatletter
\renewcommand\normalsize{%
   \@setfontsize\normalsize{9pt}{10.5pt}} 
\makeatother
\let\OLDthebibliography\thebibliography
\renewcommand\thebibliography[1]{
  \OLDthebibliography{#1}
  \setlength{\parskip}{0.5pt}
  \setlength{\itemsep}{1pt plus 0.3ex}
}
\title{Efficient and Reliable Sound Bubble Extraction for Hearables}

\name{\shortstack{Shulin He$^{1}$, Tongtao Ling$^{1}$, Pengjie Shen$^{1,2}$, Yuxuan Ma$^{1}$, Yue Qiao$^{3}$, Fan Fan$^{3}$, and Zhong-Qiu Wang$^{1}$}
}
\address{
$^{1}$Southern University of Science and Technology, Shenzhen, China\\
$^{2}$Inner Mongolia University, Hohhot, China \\
$^{3}$Institute of Media Technology and Experience Design, Huawei, Shenzhen, China \\
{
\small\texttt{\{he.shulin96, wang.zhongqiu41\}@gmail.com}}
}

\makeatletter
\providecommand{\bstctlcite}[1]{%
  \@bsphack
  \if@filesw\immediate\write\@auxout{\string\citation{#1}}\fi
  \@esphack}
\makeatother

\begin{document}
\bstctlcite{IEEEAuthorControl}
\maketitle
\begin{abstract}
In multi-talker conditions, listeners have limited ability to attend to speakers based on distance.
Recent hearable devices with sound bubble can preserve speech inside a user-specified spatial radius while suppressing speech outside.
However, these systems have high computational cost and can mistakenly suppress speech inside the selected radius or retain speech outside.
Since even occasional failures disrupt conversational continuity, improving robustness is as important as optimizing average extraction performance. 
To address this, we propose radius-conditioned TF-SkiMNet (RC-TF-SkiMNet), a lightweight causal deep neural network combining spatial features with a continuous radius input.
To strengthen radius-based selection, we train the model with two radius settings for the same mixture and propose a contrastive loss to encourage similar model outputs when both radii select the same speech, and distinct outputs otherwise.
This training incurs no computation cost at run time.
RC-TF-SkiMNet requires only $0.13$ M parameters and $1.05$ GMACs/s.
Compared with a lightweight baseline at a similar computational cost, it reduces low-scoring cases on real recordings and improves average speech quality and outside-speech suppression.

\end{abstract}

\section{Introduction}

Selective hearing is important for speech communication in cocktail-party scenarios, where concurrent speech and ambient noises can make it difficult to focus on a targeted speaker \cite{E.Cherry1953,veluri2024look}.
Existing target speech extraction (TSE) methods typically use speaker identity \cite{Zmolikova2019SpeakerBeam,Wang2018kVoiceFilter,He2020Speakerfilter,Xu2020SpEx,Hao2024XTFGridNet,shen2025listen,zeng2025usef,ling2026mc} or direction \cite{Gu2019NSF,Tesch2024SSF,zhang2025doa,li2026flexible} to select the desired speech.
However, everyday conversations often call for retaining nearby speech while suppressing speech beyond a specified distance, and using speaker identity and direction cues cannot enable this listening experience.
Sound Bubble \cite{Chen2024SoundBubble} addresses this problem by introducing radius-conditioned TSE, enabling users to preserve speech inside a specified spatial radius while suppressing speech outside.
It supports discrete and continuous radius control in real time on hearable devices.

Despite the remarkable progress made by Sound Bubble \cite{Chen2024SoundBubble}, deploying radius-conditioned TSE on resource-constrained devices still faces two challenges: computational cost and selection reliability.
To be specific, our reproduced $8$-microphone implementation of the Sound Bubble baseline \cite{Chen2024SoundBubble} requires $8.24$ GMACs/s, far exceeding the processing budget of typical
headsets.
Moreover, the system must reliably retain desired speech and suppress outside speech, as even occasional errors can disrupt user experiences.

To address the challenges, we propose radius-conditioned TF-SkiMNet (RC-TF-SkiMNet), a lightweight causal deep neural network (DNN) for hearables.
It supports continuous-radius speech extraction using only $0.13$ M parameters and $1.05$ GMACs/s, obtaining strong speech extraction performance on our hearable prototype.

In Sound Bubble \cite{Chen2024SoundBubble}, costly TF-GridNet \cite{Wang2022GridNet,Wang2023TFGridNet} is employed as the DNN backbone for TSE.
To reduce compuation, we adapt TF-SkiMNet \cite{li2025tf}, a lightweight real-time speech enhancement model, to radius-conditioned TSE by adding spatial features \cite{Wang2018combiningspectralspatial, Tan2022NSF} and a continuous radius embedding.
To further reduce computation, we compress the number of frequency bands before the recurrent blocks \cite{Rong2024GTCRN}, and replace the original global temporal memory with a causal LSTM that maintains separate states for each frequency band and shares weights across bands to limit the number of parameters.

\begin{figure}[t]
\centering
\includegraphics[width=\columnwidth]{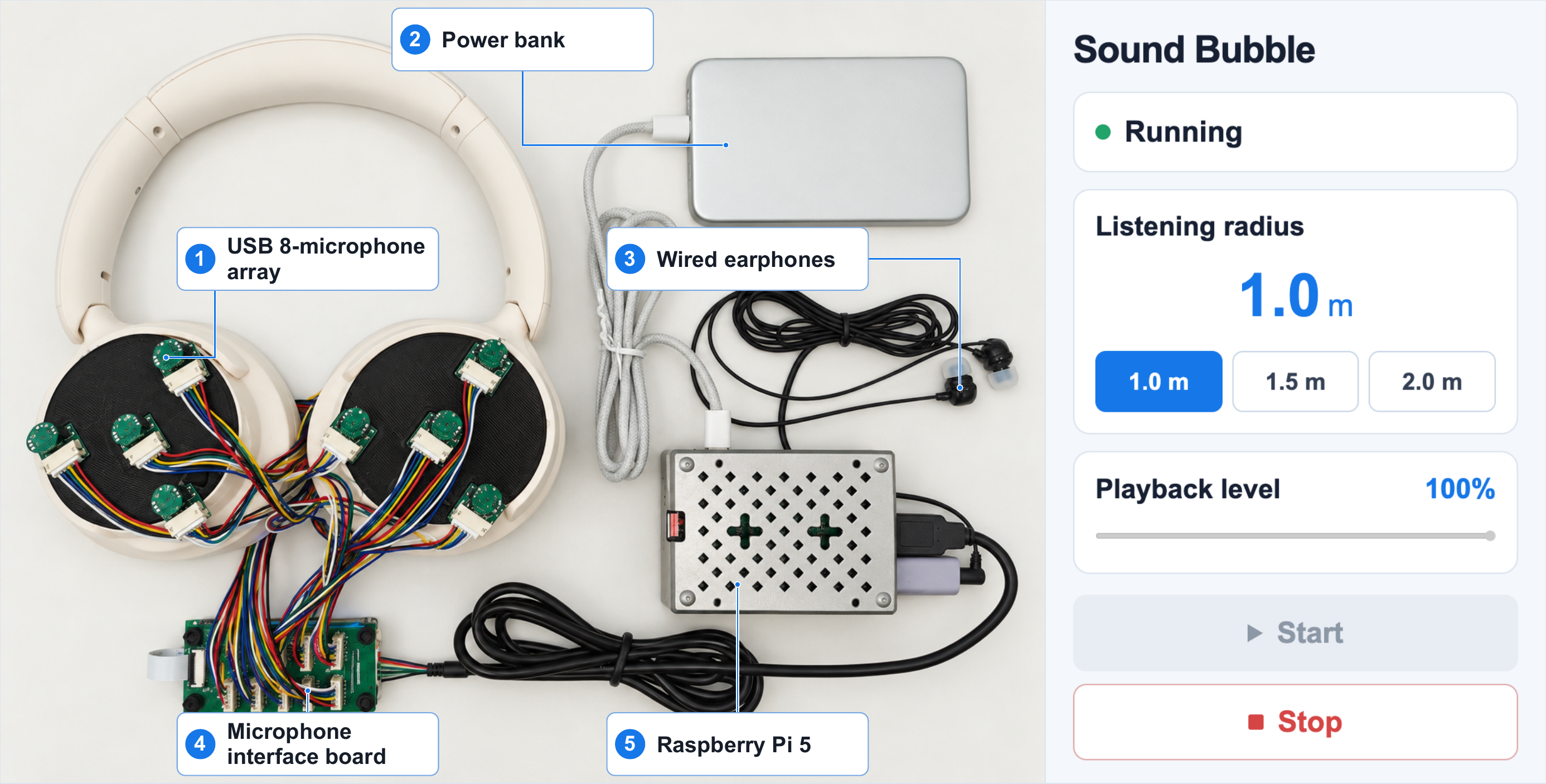}
\vspace{-0.4cm}
\caption{Prototype device and web-based control interface.}\vspace{-0.6cm}
\label{fig:prototype}
\end{figure}

To reduce extraction errors, we propose radius-paired contrastive training (RPCT) based on an insight: for a mixture, the desired speech changes only when the radius includes or excludes a speaker.
We hence feedforward the same mixture but with two different radii during training.
Besides supervising each output with its target, we introduce a contrastive loss to encourage similar projected features (calculated from the DNN outputs) when both radii select the same speech and less similar features otherwise.
Together, these losses can supervise what speech to retain and indicate how selection should change across radii.
In addition, we propose to supervise target activity and use tail losses to improve low-scoring, difficult mixtures.
Our contributions are:
\begin{itemize}[nosep, leftmargin=*]
\item We propose RC-TF-SkiMNet, a lightweight, causal, real-time DNN that supports continuous-radius TSE for an $8$-microphone headset at low inference cost.
\item We propose RPCT, which combines supervision at two radius settings with feature contrastitve learning to strengthen radius-based selection without increasing inference cost.
\end{itemize}
Evaluation results based on our headset prototype (shown in Fig. \ref{fig:prototype}) confirm the effectiveness of the proposed algorithms.
See our sound demo at {\href{https://heshulin.github.io/rc-tf-skimnet-demo/}{https://heshulin.github.io/rc-tf-skimnet-demo/}}.

\section{RC-TF-SKIMNET}\label{sec:network}

RC-TF-SkiMNet, shown in Fig. \ref{fig:network}, takes as input a radius query $d$ and an $M$-microphone signal $\mathbf x=\{x_m\}_{m=0}^{M-1}$ recorded by a headset, and estimates the speech of the speakers located inside the radius.
Let $r_k$ be the distance of speaker $k$ to the headset.
The query $d$ selects the set of speakers $\mathcal I(d)=\{k:r_k\leq d\}$.
We assume that the number of the targeted speakers is at most one,
and define target speech as $y_d = \sum\nolimits_{k\in\mathcal I(d)}s_{0,k}$, where $s_{0,k}$ is the direct-path speech of speaker $k$ at the reference microphone $0$.
Our goal is to train a DNN to estimate the target speech: $\hat y_d=\mathcal F_\theta(\mathbf x,d)$, where $\theta$ denotes the model parameters.
This section describes the input features, DNN architectures, loss functions, and training strategies.

\subsection{Input Features}

\begin{figure}[t]
\centering
\includegraphics[width=0.8\columnwidth]
{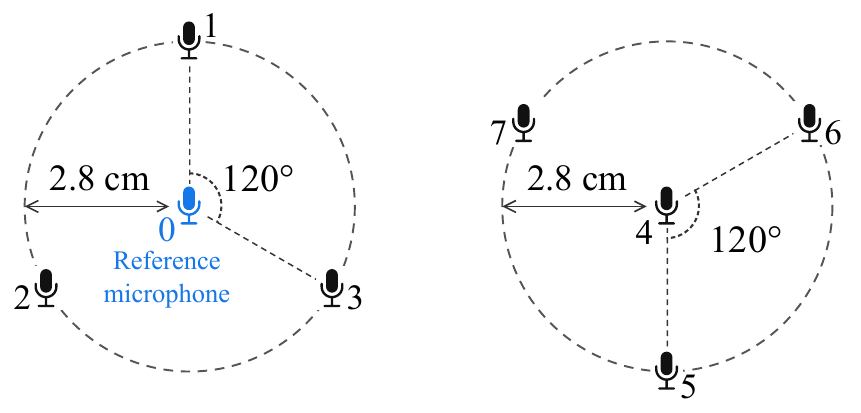}
\vspace{-0.3cm}
\caption{Microphone-array geometry of left and right earcups.}\vspace{-0.6cm}
\label{fig:layout}
\end{figure}

The input consists of a radius query and the microphone signals.
For the radius query, we asssume $d\in[0.8,2]$ m, and use \mbox{$(d-0.8)/1.2$} to map it to $[0,1]$ before further processing.
For the microphone signals, our headset uses $M=8$ microphones arranged on two earcups (see Fig. \ref{fig:layout}), each with a center microphone and $3$ outer microphones uniformly-spaced on a circle with $2.8$ cm radius.
Microphone $0$ is designated as the reference microphone.
Given the mixtures' short-time Fourier transform (STFT) coefficients $X_m(\tau,f)$, where $t$, $f$ and $m$ respectively index time frames, frequency bins and microphones,
we compute four input features: $8$-channel real and imaginary (RI) components \cite{Wang2020css}, magnitude at the reference microphone $0$, and inter-channel level and phase differences (ILD/IPD)\cite{Wang2018combiningspectralspatial, Tan2022NSF}.
For an ordered microphone pair $(i,j)$, we compute $\mathrm{ILD}_{ij}
=\log_{10}({|X_i|}/{|X_j|})$ and $\mathrm{IPD}_{ij}
=\angle X_i-\angle X_j$.
We use the sine and cosine of each IPD as input features.
The IPD pair set is $\mathcal E_{\mathrm{IPD}}=\{(1,4),\allowbreak (2,4),\allowbreak (3,4),\allowbreak (5,0),\allowbreak (6,0),\allowbreak (7,0),\allowbreak (0,4)\}$.
The ILD pair set is $\mathcal E_{\mathrm{ILD}}=\mathcal E_{\mathrm{IPD}}\cup\allowbreak\{(1,0),\allowbreak (2,0),\allowbreak (3,0),\allowbreak (5,4),\allowbreak (6,4),\allowbreak (7,4)\}$.

\subsection{DNN Architecture}

\begin{figure}[t]
\centering
\includegraphics[width=1.00\columnwidth]
{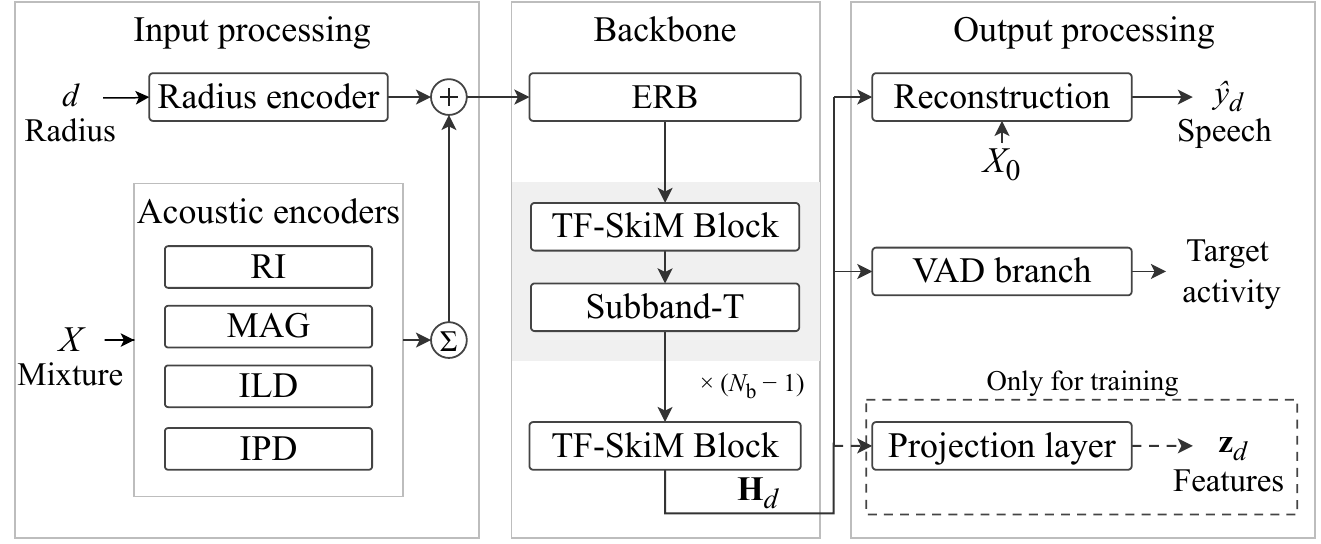}
\vspace{-0.4cm}
\caption{
Architecture of RC-TF-SkiMNet.
}\vspace{-0.4cm}
\label{fig:network}
\end{figure}

As shown in Fig. \ref{fig:network}, RC-TF-SKiMNet takes microphone signals $\mathbf{x}$ and a radius query $d$ as inputs, and models time-frequency (T-F) patterns in compressed frequency bands to reconstruct target speech.

It first separately encodes the $4$ input features (i.e., RI, reference magnitude, ILD and IPD).
Each encoder uses a causal temporal convolution, layer normalization (LN), and a parametric rectified linear unit (PReLU) to produce $C$ hidden channels.
We summate the $4$ $C$-channel outputs to combine the acoustic information without increasing the channel dimension.
To guide the network to extract speech within the requested radius, we encode the normalized radius query using $2$ linear layers with $C_r$ and $C$ output units, each followed by LN and PReLU.
We then add the radius embedding to the combined acoustic features at every T-F bin.

Next, we compress the $F$ frequency bands to reduce computation.
Following GTCRN \cite{Rong2024GTCRN}, we use a fixed equivalent rectangular bandwidth (ERB) transform \cite{Glasberg1990ERB}.
It retains $F_l$ low-frequency bands and maps the remaining $F-F_l$ bands to $F_h$ bands through a fixed triangular matrix $\mathbf W_{\mathrm{ERB}}\in\mathbb R^{F_h\times(F-F_l)}$.
This reduces the number of frequency bands to model from $F$ to $F_l+F_h$.

We use $N_b$ processing blocks to model T-F patterns in the compressed features.
Each block follows the TF-SkiMNet structure \cite{li2025tf}: \textit{Local-F} captures local spectral patterns through frequency convolution, \textit{Full-band F} combines information across frqeuency bands through a bidirectional long short-term memory (LSTM) layer within each frame, and \textit{Local-T} captures short-term changes through causal temporal convolution.
To capture longer-term patterns in speech, we use a \textit{Subband-T} module after each of the first $N_b-1$ blocks, replacing TF-SkiMNet's global time memory.
\textit{Subband-T} uses a causal LSTM to maintain a separate temporal state for each band, while sharing weights across bands to limit the number of parameters \cite{Hao2021FullSubNet}.
All modules use residual connections, and both the hidden channel dimension and LSTM hidden size are set to $C$.
The final block outputs $\mathbf H_d$, which is fed to $3$ parallel branches: speech reconstruction, voice activity detection (VAD), and feature projection for RPCT.

The reconstruction branch restores the original frequency resolution by expanding the compressed bands with $(\mathbf W_{\mathrm{ERB}})^\top$.
A causal decoder then predicts the RI components of a complex mask \cite{Williamson2016}, each bounded by $\tanh$ to the range $[-1,1]$.
Applying this mask to the reference STFT spectrogram $X_0$ and taking inverse STFT yields estimated target speech $\hat y_d$.

The VAD branch averages $\mathbf H_d$ over frequency to obtain a $C$-dimensional embedding per frame.
LN, a linear layer with $C$ output units, PReLU, and a scalar output layer then produce a target activity score for each frame.

For RPCT, the projection layer first averages $\mathbf H_d$ over time and frequency to obtain a single embedding.
After LN, the embedding passes through two linear layers with $D_p$ output units each and a PReLU between them.
$L_2$ normalization then produces a unit-length vector $\mathbf z_d\in\mathbb R^{D_p}$ for contrastive learning.
Note that this branch is used only during training.

\subsection{Training Strategies and Loss Functions}
\label{sec:paired}

\begin{figure}[t]
\centering
\includegraphics[width=0.8\columnwidth]
{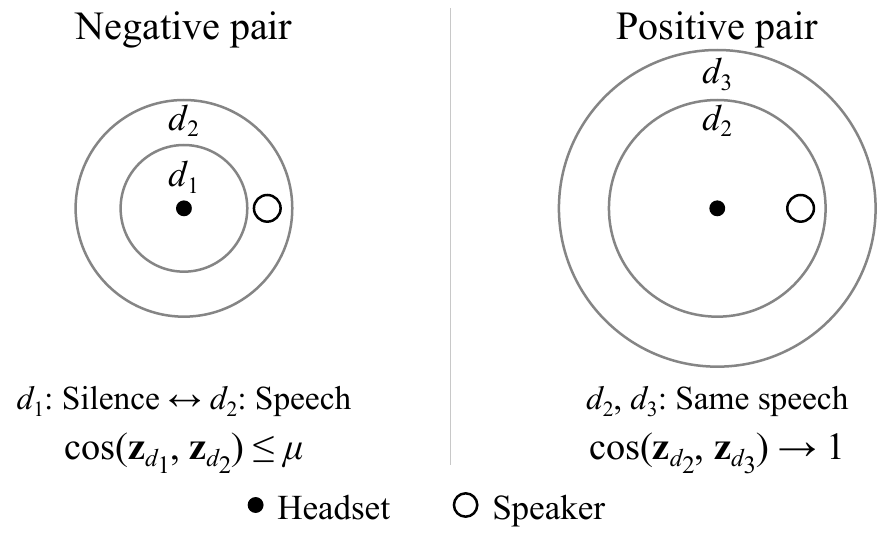}
\vspace{-0.3cm}
\caption{
Illustration of RPCT.}\vspace{-0.6cm}
\label{fig:radius_pairs}
\end{figure}

To reduce speech selection errors,  we train the network with target reconstruction, VAD, RPCT loss, and tail losses.

For each radius query, we supervise the estimated target speech using waveform and spectral losses:
\begin{equation}
\mathcal L_{\mathrm{SEP}}
=\mathcal L_{\mathrm{SNR}}
+\mathcal L_{\mathrm{RI}}
+\mathcal L_{\mathrm{MAG}},
\label{eq:sep_loss}
\end{equation}
where $\mathcal L_{\mathrm{SNR}}$ is the time-domain negative SNR loss \cite{le2019sdr}, 
$\mathcal L_{\mathrm{RI}}$ an $L_1$ loss defined on power-compressed real and imaginary components, and $\mathcal L_{\mathrm{MAG}}$ an $L_1$ loss on power-compressed magnitudes \cite{Wang2020chime,Braun2021Losses}.

The VAD loss adds explicit activity supervision to target reconstruction.
We train the VAD branch with frame-level binary cross-entropy loss $\mathcal L_{\mathrm{VAD}}$ \cite{lin2021sparsely}, encouraging the shared features $\mathbf H_d$ to distinguish frames with and without target speech.

RPCT extends the per-query supervision to a radius-pair settings.
For a training mixture, the target signal changes only when the radius includes or excludes a speaker.
We thus pair a main query $d$ with a second query $d'$ of the same mixture.
Under our single-target-speaker assumption, positive pairs select the same speaker or both select no speaker, while negative pairs contain one target-present query and one target-absent query.

We compare the corresponding normalized features $\mathbf z_d$ and $\mathbf z_{d'}$ from the shared projection head (see Fig. \ref{fig:radius_pairs}).
Based on their cosine similarity $c=\mathbf z_d^\top\mathbf z_{d'}$, we define the contrastive loss \cite{Hadsell2006Contrastive} as
\begin{equation}
\mathcal L_{\mathrm{RPCT}}(d,d')=
\begin{cases}
1-c, & \mathcal I(d)=\mathcal I(d'),\\
[\max(0,c-\mu)]^2, & \mathcal I(d)\ne\mathcal I(d'),
\end{cases}
\label{eq:contrastive}
\end{equation}
where positive pairs are encouraged to have similar features, while negative ones are penalized only when $c$ is larger than $\mu$, a pre-defined similarity margin.

The tail losses give additional weight to main queries with poor target recovery or suppression.
We score target-present queries using scale-invariant signal-to-distortion ratio improvement (SI-SDRi) \cite{le2019sdr} and target-absent queries using the energy suppression ratio (ESR), both expressed in dB:
\begin{equation}
q_d=
\begin{cases}
\operatorname{SI-SDR}(\hat y_d,y_d)
-\operatorname{SI-SDR}(x_0,y_d), & y_d\ne\mathbf{0},\\[3pt]
10\log_{10}\!\bigl[(\|x_0\|_2^2+\epsilon)/
(\|\hat y_d\|_2^2+\epsilon)\bigr], & y_d=\mathbf{0}.
\end{cases}
\label{eq:tail_score}
\end{equation}
Following conditional value-at-risk \cite{Rockafellar2000CVaR}, each tail loss averages the largest fraction $\rho$ of the penalties $\ln(1+\exp(\kappa-q_d))$ within its target group.
Here, $\kappa$ is the group-specific score threshold, and both $\kappa$ and $q_d$ are expressed in dB.
We denote the target-present and target-absent tail losses by $\mathcal T_a$ and $\mathcal T_n$, respectively.

The overall loss is defined as:
\begin{equation}
\mathcal L_{\mathrm{FULL}} = \mathcal L_{\mathrm{SEP}}
+\lambda_v\mathcal L_{\mathrm{VAD}}
+\lambda_a\mathcal T_a
+\lambda_n\mathcal T_n
+\lambda_p\mathcal L_{\mathrm{RPCT}},
\label{eq:full_loss}
\end{equation}
where $\lambda_v$, $\lambda_a$, $\lambda_n$ and $\lambda_p$ are tunable weighting terms.

\section{Experimental Setup}
\subsection{Data and Reference Signals}\label{sec:data}

Unless otherwise noted, we train all systems on simulated mixtures and remixed single-speaker headset recordings using a co-learning strategy \cite{wang2025superm2m}.
The sampling rate is $16$ kHz.

For simulated mixtures, clean speech signals are sampled from LibriSpeech \cite{panayotov2015librispeech}, where \textit{train-clean-\{100,360\}} are used for training, \textit{dev-clean} for validation, and \textit{test-clean} for testing.
To make the simulated data reflect the acoustic effects of head shadowing, we simulate headset responses with COMSOL \cite{COMSOLMultiphysics} and generate $8$-channel room impulse responses (RIRs) using the image method \cite{Allen1979Image}.
Among the simulated scenes, $80\%$ have $T_{60}\in[0.3,0.8]$ s, while the remaining $20\%$ have $T_{60}\in(0.8,1.2]$ s.
We convolve each clean speech signal with its corresponding RIRs to generate multi-channel mixtures.
The target speaker's direct-path speech at microphone $0$ is used as the reference.
For background noise, we use two-channel recordings from WHAM! \cite{Wichern2019}.
Instead of convolving the noise with RIRs, we add its two channels directly to the microphones on the left and right earcups, respectively, so that all the microphones on an earcup share the same noise signal.
We refer to this as ear-level noise, and add it to half of the simulated mixtures.
For training and validation mixtures containing both target and interference speakers, we sample the target-to-interference energy ratio uniformly from the range of $[-5,10]$ dB.
Training uses $20,000$ scenes, while validation uses simulated mixtures from $2,000$ different scenes.
For testing, we generate $500$ mixtures from $500$ held-out scenes and an additional $100$ single-speaker mixtures for radius-sweeping evaluation.

To adapt the model to real signals, we collect single-speaker recordings using our headset prototype (see Fig. \ref{fig:prototype}) worn by $10$ participants.
Recordings collected with $5$ participants are used for training, and the ones with the remaining $5$ are used for testing.
We remix the raw multi-channel single-speaker recordings to generate training mixtures.
We use a pre-trained monuaral TF-SkiMNet \cite{li2025tf} to denoise the reference microphone
to obtain a single-channel target reference signal.
The recordings will be released, and collection details are available on our demo page\footnote{See \href{https://heshulin.github.io/rc-tf-skimnet-demo/\#recording-protocol}{https://heshulin.github.io/rc-tf-skimnet-demo/\#recording-protocol}.}.

\subsection{Baselines and System Configurations}

We compare the systems in Table \ref{tab:systems} to evaluate the lightweight extraction design and training strategy. M1 is a small TF-GridNet-based Sound Bubble baseline \cite{Chen2024SoundBubble,Wang2023TFGridNet}, and M2 is RC-TF-SkiMNet trained only with reconstruction loss $\mathcal L_{\mathrm{SEP}}$. We set their LSTM hidden sizes respectively to $23$ and $22$, to compare the designs at similar parameter counts and computational cost. M1 retains Sound Bubble's complex-spectrum prediction, while M2 uses the complex-mask reconstruction described in Section \ref{sec:network}. M3 adds VAD, tail losses, and RPCT to M2, evaluating the combined training strategy with the same architecture. M1 and M3 provide the main comparison between the baseline and our complete system.

M1 uses the Sound Bubble training loss \cite{Chen2024SoundBubble}, denoted by Base in Table \ref{tab:systems}:
\begin{equation}
\mathcal L_{\mathrm{BASE}}=
\begin{cases}
\mathcal L_{\mathrm{SNR}}(\hat y_d,y_d), & y_d\ne\mathbf{0},\\[2pt]
\displaystyle50\times \frac{1}{T}\sum_{t=1}^{T}|\hat y_d(t)|, & y_d=\mathbf{0},
\end{cases}
\label{eq:base_loss}
\end{equation}
where $\mathcal L_{\mathrm{SNR}}$ is the negative waveform SNR loss and $T$ the number of waveform samples. The second term penalizes residual output when no target is selected.

All systems use $6$ blocks and a $320$-point STFT with a $20$ ms square-root Hann window and a $10$ ms hop.

For RC-TF-SkiMNet, we set $F=161$, $F_l=30$, $F_h=54$,
$C_r=16$, and $D_p=64$ (RPCT).
The kernel sizes for all the encoders, as well as the \textit{Local-T} and \textit{Local-F} modules, are set to $5$.

For reconstruction, we use a magnitude compression exponent of $0.3$ and normalize the spectral losses per example by the summated compressed input magnitude at the reference microphone.
 
For RPCT, we only apply it on simulated mixtures.
Both query radii lie within $[0.8,2]$ m, and each query selects at most one target speaker.
We set the negative-pair similarity margin $\mu$ to $0.2$.

For the tail losses, we set $\rho=0.25$ and score thresholds of $6$ dB for target-present queries and $15$ dB for target-absent queries.
ESR is computed with $\epsilon=10^{-8}$.
The loss weights in Eq. (\ref{eq:full_loss}) are $\lambda_v=0.1$, $\lambda_a=\lambda_n=0.05$, and $\lambda_p=0.25$.

\begin{table}[t]
\centering
\scriptsize
\caption{System configurations, where (1) $C$ denotes hidden channels and $H$ LSTM hidden size; (2) all systems use continuous radius queries in $[0.8,2]$ m; (3) ``Factor $3$'' uses a frequency-axis convolution (kernel size and stride $3$) before each frequency LSTM, followed by transposed convolution to restore resolution \cite{Chen2024SoundBubble}; (4) ``ERB'' uses the fixed mapping in Section \ref{sec:network}; and (5) $\mathcal L_{\mathrm{BASE}}$ is defined in Eq. \eqref{eq:base_loss}, while $\mathcal L_{\mathrm{FULL}}$ is defined in Eq. \eqref{eq:full_loss}.}
\vspace{-0.1cm}
\label{tab:systems}
\setlength{\tabcolsep}{2pt}
\begin{threeparttable}
\begin{tabular*}{\columnwidth}{@{\extracolsep{\fill}}lcccl@{}}
\toprule
System & $C$ & $H$ & Frequency Compression & Loss\\
\midrule
M1: TF-GridNet\cite{Chen2024SoundBubble} & $23$ & $23$ & Factor $3$ & $\mathcal L_{\mathrm{BASE}}$\\
M2: RC-TF-SkiMNet & $22$ & $22$ & ERB & $\mathcal L_{\mathrm{SEP}}$\\
M3: RC-TF-SkiMNet & $22$ & $22$ & ERB & $\mathcal L_{\mathrm{FULL}}$\\
\bottomrule
\end{tabular*}
\end{threeparttable}
\vspace{-0.3cm}
\end{table}

\section{Evaluation Results}
\label{sec:Results}

For evaluation, we consider two types of data: simulated and recorded mixtures.
The query radii are $1$, $1.5$, and $2$ m for simulated mixtures and $1$ m for recorded mixtures.
In both data types, we evaluate TSE under four test conditions: one target speaker with 1) a background noise source; 2) an interference speaker; 3) an interference speaker plus a noise source; and 4) two interference speakers.
Each condition contains $100$ test mixtures per data type.
We report mean SI-SDRi \cite{le2019sdr} and wideband perceptual evaluation of speech quality (PESQ-WB) \cite{rix2001perceptual,ITU2007P8622} over all target-present mixtures, separately for each data type.
For $100$ target-absent mixtures per domain, we report the mean ESR in dB, which is defined in Eq. \eqref{eq:tail_score}.
To assess extraction reliability, we report $F_{\mathrm{SI}}$, defined as the percentage of target-present mixtures with SI-SDRi less than $6$ dB, and $F_{\mathrm{ESR}}$, defined as the percentage of target-absent mixtures with ESR less than $15$ dB.

\begin{table}[t]
\centering
\scriptsize
\captionsetup{skip=3pt}
\caption{Inference cost and extraction performance on simulated (Sim.) and recorded (Rec.) mixtures. Bold: best among M1--M3 for each data type.}
\vspace{-0.1cm}
\label{tab:results}
\setlength{\tabcolsep}{1.2pt}
\begin{tabular*}{\columnwidth}{
@{\extracolsep{\fill}}
l
c
c
c
S[table-format=2.1,round-precision=1]
S[table-format=1.2]
S[table-format=2.1,round-precision=1]
S[table-format=2.1,round-precision=1]
S[table-format=2.1,round-precision=1]@{}
}
\toprule
& & & & \multicolumn{3}{c}{Average Performance}
& \multicolumn{2}{c}{Failure Rate}\\
\cmidrule(lr){5-7}\cmidrule(lr){8-9}
System
& \makecell{\#params\\(M)}
& \makecell{Compute\\(GMACs/s)}
& Data
& \multicolumn{1}{c}{\makecell{SI-SDRi\\(dB)$\uparrow$}}
& \multicolumn{1}{c}{\makecell{PESQ-\\WB$\uparrow$}}
& \multicolumn{1}{c}{\makecell{ESR\\(dB)$\uparrow$}}
& \multicolumn{1}{c}{\makecell{$F_{\mathrm{SI}}$\\(\%)$\downarrow$}}
& \multicolumn{1}{c}{\makecell{$F_{\mathrm{ESR}}$\\(\%)$\downarrow$}}\\
\midrule
\multirow{2}{*}{M1}
& \multirow{2}{*}{\bfseries\tablenum[table-format=1.2]{0.13}}
& \multirow{2}{*}{\tablenum[table-format=1.2]{1.09}}
& Sim. & \bfseries 9.22 & \bfseries 2.00 & 32.76 & 17.00 & 24.00\\
& & & Rec. & 11.83 & 2.15 & 36.89 & 1.00 & 8.00\\
[0.5ex]\hdashline\noalign{\vskip 0.5ex}
\multirow{2}{*}{M2}
& \multirow{2}{*}{\bfseries\tablenum[table-format=1.2]{0.13}}
& \multirow{2}{*}{\bfseries\tablenum[table-format=1.2]{1.05}}
& Sim. & 9.04 & 1.96 & 52.80 & 16.50 & 18.00\\
& & & Rec. & 11.49 & 2.15 & \bfseries 68.85 & 7.50 & 3.00\\
[0.5ex]\hdashline\noalign{\vskip 0.5ex}
\multirow{2}{*}{M3}
& \multirow{2}{*}{\bfseries\tablenum[table-format=1.2]{0.13}}
& \multirow{2}{*}{\bfseries\tablenum[table-format=1.2]{1.05}}
& Sim. & 9.18 & 1.94 & \bfseries 57.14 & \bfseries 15.75 & \bfseries 16.00\\
& & & Rec. & \bfseries 12.50 & \bfseries 2.16 & 68.53 & \bfseries 0.50 & \bfseries 1.00\\
\bottomrule
\end{tabular*}
\vspace{-0.3cm}
\end{table}

Table \ref{tab:results} compares extraction performance and inference cost.
Reconstruction-only M2 provides comparable mean SI-SDRi and PESQ-WB and higher ESR than M1 for both simulated and recorded mixtures, but has a higher $F_{\mathrm{SI}}$ for recorded mixtures.
With the complete training strategy, on recorded mixtures M3 reduces $F_{\mathrm{SI}}$ from $7.50\%$ to $0.50\%$ without increasing inference cost. 
M3 outperforms M1 on recorded mixtures, raising SI-SDRi from $11.8$ to $12.5$ dB and reducing $F_{\mathrm{ESR}}$ from $8.0\%$ to $1.0\%$, while lowering both failure rates on simulated mixtures.
As shown in Table \ref{tab:results}, these gains require only $0.13$M parameters and $1.05$ GMACs/s, compared with $1.09$ GMACs/s for M1.

Table \ref{tab:ablations} reports ablation results on the training losses, frequency compression, input features, and training data.
Starting from M2, VAD supervision reduces $F_{\mathrm{SI}}$ on recorded mixtures from $7.50\%$ to $1.75\%$, but raises $F_{\mathrm{ESR}}$ on simulated mixtures from $18.0\%$ to $29.0\%$.
The tail losses reduce the latter to $16.0\%$, at the cost of a lower SI-SDRi on simulated mixtures ($8.4$ vs. $9.6$ dB).
RPCT then restores this SI-SDRi to $9.2$ dB while maintaining the same $F_{\mathrm{ESR}}$, and further reduces $F_{\mathrm{SI}}$ on recorded mixtures from $2.00\%$ to $0.50\%$ at a similar mean SI-SDRi.
The three training components are thus complementary in improving extraction reliability.
Next, we remove ERB compression, ILD, or IPD from M3.
Removing ERB increases computation to $1.87$ GMACs/s, lowers the SI-SDRi for both data types, and raises $F_{\mathrm{ESR}}$ on recorded mixtures from $1.0\%$ to $12.0\%$.
Removing ILD or IPD raises $F_{\mathrm{SI}}$ on recorded mixtures from $0.50\%$ to $4.75\%$ or $1.25\%$, and $F_{\mathrm{ESR}}$ on simulated mixtures from $16.0\%$ to $30.0\%$ or $21.0\%$, respectively.
These results support the use of ERB compression for computationally efficient extraction, and the joint use of ILD and IPD for reliable speech selection.
Finally, training M3 only on simulated mixtures raises the SI-SDRi on simulated mixtures to $10.4$ dB but reduces that on recorded mixtures to $1.2$ dB, demonstrating the importance of recorded training data.

Table \ref{tab:continuous} reports performance across query radii.
We query $100$ single-speaker mixtures per domain at $1$, $1.2$, $1.4$, $1.6$, $1.8$, and $2$ m.
Source distances range from $0.5$ to $2.5$ m in simulation and are fixed at $1.75$ m for recordings.
PESQ-WB and eSTOI \cite{Jensen2016ESTOI} are averaged over inside-radius queries, and ESR over outside-radius queries.
M3 improves all metrics over M1 for both data types, including an eSTOI of $0.742$ vs. $0.379$ and an ESR of $50.6$ vs. $33.5$ dB on recorded mixtures.
These results demonstrate the stronger radius-based speech selection capability of RC-TF-SkiMNet.
Overall, the RC-TF-SkiMNet keeps inference cost low, while RPCT improves extraction reliability, making continuous-radius TSE more practical for hearables.

\begin{table}[!t]
\centering
\scriptsize
\captionsetup{skip=3pt}
\caption{Ablation results based on test sets.}
\vspace{-0.1cm}
\label{tab:ablations}
\setlength{\tabcolsep}{1.2pt}
\begin{tabular*}{\columnwidth}{
@{\extracolsep{\fill}}
l
S[table-format=2.1,round-precision=1]
S[table-format=2.1,round-precision=1]
S[table-format=2.1,round-precision=1]
S[table-format=2.2]
S[table-format=2.1,round-precision=1]
@{}
}
\toprule
& \multicolumn{2}{c}{Simulated}
& \multicolumn{3}{c}{Recorded}\\
\cmidrule(lr){2-3}\cmidrule(lr){4-6}
System
& \multicolumn{1}{c}{\makecell{SI-SDRi\\(dB)$\uparrow$}}
& \multicolumn{1}{c}{\makecell{$F_{\mathrm{ESR}}$\\(\%)$\downarrow$}}
& \multicolumn{1}{c}{\makecell{SI-SDRi\\(dB)$\uparrow$}}
& \multicolumn{1}{c}{\makecell{$F_{\mathrm{SI}}$\\(\%)$\downarrow$}}
& \multicolumn{1}{c}{\makecell{$F_{\mathrm{ESR}}$\\(\%)$\downarrow$}}\\
\midrule
\multicolumn{6}{l}{Losses}\\
Separation (M2)       & 9.04 & 18.00 & 11.49 & 7.50 & 3.00\\
\quad $+$ VAD         & 9.62 & 29.00 & 12.52 & 1.75 & 1.00\\
\quad $+$ Tail losses & 8.40 & 16.00 & 12.47 & 2.00 & 1.00\\
\quad $+$ RPCT (M3)   & 9.18 & 16.00 & 12.50 & 0.50 & 1.00\\
\midrule
\multicolumn{6}{l}{Features (removed from M3)}\\
\quad $-$ ERB         & 8.99 & 17.00 & 11.93 & 3.25 & 12.00\\
\quad $-$ ILD and IPD & 9.03 & 30.00 & 11.66 & 3.25 & 0.00\\
\quad $-$ IPD         & 9.60 & 21.00 & 12.32 & 1.25 & 0.00\\
\quad $-$ ILD         & 8.48 & 30.00 & 11.18 & 4.75 & 2.00\\
\midrule
\multicolumn{6}{l}{Training data (M3 variant)}\\
Simulation only & 10.36 & 21.00 & 1.22 & 77.50 & 9.00\\
\bottomrule
\end{tabular*}
\vspace{-0.3cm}
\end{table}

\begin{table}[!t]
\centering
\scriptsize
\captionsetup{skip=3pt}
\caption{Radius-sweeping results on simulated and recorded mixtures.
}
\vspace{-0.1cm}
\label{tab:continuous}
\setlength{\tabcolsep}{1.2pt}
\begin{tabular*}{\columnwidth}{
@{\extracolsep{\fill}}
l
S[table-format=1.2]
S[table-format=1.3,round-precision=3]
S[table-format=2.1,round-precision=1]
S[table-format=1.2]
S[table-format=1.3,round-precision=3]
S[table-format=2.1,round-precision=1]
@{}
}
\toprule
& \multicolumn{3}{c}{Simulated}
& \multicolumn{3}{c}{Recorded}\\
\cmidrule(lr){2-4}\cmidrule(lr){5-7}
System
& \multicolumn{1}{c}{\makecell{PESQ-WB$\uparrow$}}
& \multicolumn{1}{c}{eSTOI$\uparrow$}
& \multicolumn{1}{c}{\makecell{ESR (dB)$\uparrow$}}
& \multicolumn{1}{c}{\makecell{PESQ-WB$\uparrow$}}
& \multicolumn{1}{c}{eSTOI$\uparrow$}
& \multicolumn{1}{c}{\makecell{ESR (dB)$\uparrow$}}\\
\midrule
M1 & 2.77 & 0.894 & 24.92 & 1.29 & 0.379 & 33.45\\
M3 & \bfseries 2.90 & \bfseries 0.916 & \bfseries 47.92 & \bfseries 1.64 & \bfseries 0.742 & \bfseries 50.57\\
\bottomrule
\end{tabular*}
\vspace{-0.3cm}
\end{table}

\section{Conclusion}

We have proposed RC-TF-SkiMNet, a lightweight causal network for continuous-radius speech extraction on hearables, and RPCT, a training strategy that supervises the same mixture at two radius settings and contrasts the resulting features.
RC-TF-SkiMNet requires only $0.13$ M parameters and $1.05$ GMACs/s, and RPCT incurs no additional inference cost.
Compared with a lightweight Sound Bubble baseline at a similar inference cost, the proposed system yields fewer low-scoring cases on recorded mixtures, and better speech quality and outside-speech suppression in radius sweeping.

\clearpage
{\footnotesize
\bibliographystyle{IEEEtran}
\bibliography{refs}
}
\end{document}